\input harvmac \newcount\figno \figno=0 \def\fig#1#2#3{
\par\begingroup\parindent=0pt\leftskip=1cm\rightskip=1cm\parindent=0pt
\baselineskip=11pt \global\advance\figno by 1 \midinsert \epsfxsize=#3
\centerline{\epsfbox{#2}} \vskip 12pt {\bf Fig. \the\figno:} #1\par
\endinsert\endgroup\par } \def\figlabel#1{\xdef#1{\the\figno}}
\def\encadremath#1{\vbox{\hrule\hbox{\vrule\kern8pt\vbox{\kern8pt
\hbox{$\displaystyle #1$}\kern8pt} \kern8pt\vrule}\hrule}}

\overfullrule=0pt

%
\def\tilde{\widetilde}
\def\bar{\overline}

\font\zfont = cmss10 

\def\bigone{\hbox{1\kern -.23em {\rm l}}}
\def\ZZ{\hbox{\zfont Z\kern-.4emZ}}

\Title{hep-th/9804188, IASSNS-HEP-98/36}
{\vbox{\centerline{ On the Deformation Quantization Description }
\centerline{of Matrix Compactifications}}}
\medskip
\centerline{Hugo
Garc\'{\i}a-Compe\'an\foot{Also {\it Departamento de
F\'{\i}sica, Centro de Investigaci\'on y de Estudios Avanzados del IPN,
Apdo. Postal 14-740, 07000, M\'exico D.F., M\'exico}. E-mail:
compean@sns.ias.edu}}
\smallskip
\centerline{\it School of Natural Sciences}
\centerline{\it Institute for Advanced Study}
\centerline{\it Olden Lane, Princeton, NJ 08540, USA}
\vskip .5truecm

\bigskip
\medskip
\vskip 1truecm
\noindent

Matrix theory compactifications on tori have associated Yang-Mills
theories on the dual tori with sixteen supercharges. A noncommutative
description of these Yang-Mills theories based in deformation quantization
theory is provided. We show that this framework allows a natural
generalization of the `Moyal $B$-deformation' of the Yang-Mills theories
to non-constant background $B$-fields on curved spaces.  This
generalization is described through Fedosov's geometry of deformation
quantization.




\Date{April, 1998}


\vfill
\break

\newsec{Introduction}

Matrix theory is by now the best candidate to realize the non-perturbative
quantum theory underlying string theory termed $M$-theory (for recent
reviews see \ref\review{ T. Banks, hep-th/9710231; D. Bigatti and L.
Susskind, hep-th/9712072; W. Taylor IV, hep-th/9801182; A. Sen,
hep-th/9802051. }).

Compactifications of matrix theory on the tori of dimension $p\leq 4$ were
shown to be equivalent to $(p+1)$-dimensional Yang-Mills theory with 16
supercharges \nref\bfss{T. Banks, W. Fischler, S.H. Shenker and L.
Susskind, ``$M$-theory as the Matrix Model:  A Conjecture'', Phys. Rev. D
{\bf 55} (1997) 5112, hep-th/9610043.} \nref\taylor{W. Taylor IV,
``D-brane Field Theory on Compact Spaces'', Phys. Lett.  B {\bf 394}
(1997) 283, hep-th/9611042.} \nref\ganor{O.J. Ganor, S. Ramgoolam and W.
Taylor IV, ``Branes, Fluxes and Duality in Matrix Theory'', Nucl. Phys. B
{\bf 492} (1997) 191, hep-th/9611202.} \refs{\bfss,\taylor,\ganor}. For
$p=5,6$ it is known that additional degrees of freedom to those coming
from $D$-branes wrapped around homology cycles of lower dimensions arise
and new prescriptions have to be considered \nref\nati{N. Seiberg, ``New
theories in Six Dimensions and Matrix Description of $M$-theory on $T^5$
and $T^5/Z_2$'', hep-th/9705221.} \nref\sen{A. Sen, ``$D$0-branes on $T^n$
and Matrix theory'', hep-th/9709220.} \nref\seiberg{ N. Seiberg, ``Why is
the Matrix Model Correct?'', Phys. Rev. Lett. {\bf 79} (1997) 3577,
hep-th/9710009.} \refs{\nati,\sen,\seiberg}. Aside from compactifications
on the torus, matrix compactifications on curved manifolds have been
considered in \nref\ooguri{M.R. Douglas, H. Ooguri and S. Shenker, Phys.
Lett. B {\bf 402} (1997) 36, hep-th/9702203.} \nref\curved{M.R. Douglas,
``D-Branes in Curved Space'', hep-th/9703056; ``$D$-branes and Matrix
Theory in Curved Space'', hep-th/9707228.} \nref\volovich{I.V. Volovich,
hep-th/9705013; I. Ya. Aref'eva and I.V. Volovich, ``Matrix Theory in
Curved Space'', hep-th/9802091.} \refs{\ooguri,\curved} and in a different
context in \refs{\volovich}. Recently compactifications on Calabi-Yau
threefolds have also considered in \ref\eva{ S. Kachru, A. Lawrence and E.
Silverstein, ``On the Matrix Description of Calabi-Yau
Compactifications'', hep-th/9712223.}. There it was shown that in DKPS
limit, Calabi-Yau compactifications are simpler than ${\bf T}^6$
compactifications of matrix theory. In the DLCQ description the remaining
degrees of freedom are decoupled from gravity.  This difference apparently
has to do with the different topological properties of the Calabi-Yau and
${\bf T}^6$ spaces.

In a seminal paper by Connes, Douglas and Schwarz the
interconnections between matrix theory and non-commutative geometry were explored
\ref\connes{A. Connes, M.R. Douglas and A. Schwarz, ``Noncommutative Geometry and
Matrix Theory: Compactification on Tori'', hep-th/9711162.}. In particular it was
shown that toroidal compactifications of matrix theory with non-vanishing
supergravity
background three form on the torus, can be described as a gauge theory
on a non-commutative dual torus of the type discussed in \ref\alan{A. Connes,
{\it Noncommutative Geometry}, Academic Press, 1994.}. The physical justification
of this relation involves the description of matrix theory in terms
of $D$-branes on  backgrounds which include a  tensor $C_{-ij}$ on the torus \ref\hull{
M. Douglas and C. Hull, ``D-Branes and Non-commutative Torus'', hep-th/9711165.}.

The study of compactification on non-commutative tori and orbifolds
of matrix theory was discussed in \ref\wu{ Pei-Ming Ho, Yi-Yen Wu and Yong-Shi Wu,
``Towards a Non-commutative Geometric Approach to Matrix Compactification'',
hep-th/9712201;  Pei-Ming Ho and Yong-Shi Wu, ``Noncommutative Gauge Theories
in Matrix Theory'', hep-th/9801147.}. In further developments an embedding of
non-commutative
compactifications of matrix theory in weak coupling string theory is discussed in
\ref\li{Miao Li, ``Comments on Supersymmetric Yang-Mills Theory on a Non-commutative
Torus'', hep-th/9802052.}.

An explicit construction of the $(p+1)$-dimensional Yang-Mills theory with 16
supercharges,
from $D0$-brane action of type IIA superstring theory in a constant $B$-field
background
was done very recently in \nref\cheung{ Y-K. E. Cheung and M. Krogh, ``Non-commutative
Geometry from 0-branes in a Background $B$-field, hep-th/9803031}
\nref\kawano{ T. Kawano and
K. Okuyama, ``Matrix Theory on Noncommutative Torus'', hep-th/9803044.}
\refs{\cheung,\kawano}. These theories are supersymmetric Yang-Mills theories
 on noncommutative
spaces and are non-local field theories of the type described in \nref\berkooz
{ M. Berkooz, ``Non-local Field Theories and
the Non-commutative Torus'', hep-th/9802069.} \berkooz.  UV fixed points do not
exist in these theories
and IR and UV behaviors are disentangled by the prescription of taking the size of
the torus infinite, keeping  the size of the non-local scale fixed.  This
prescription  works out also for (2,0) field theories in six-dimensions in the DLCQ
description. Here the noncommutativity is reflected in the resolution of ALE
singularities. The description of $D$ branes in a background $B$-field has
mathematical applications as well. An example is the ADHM construction of Yang-Mills
instantons on noncommutative spaces \ref\schwarz{N. Nekrasov and A. Schwarz,
``Instantons on Noncommutative ${\bf R}^4$, and (2,0) Superconformal
Six-dimensional Theory'', hep-th/9802068.}. (0,2) field theory newly fit into
this picture.

Thus the effect of the background $B$-field matrix compactifications on curved
spaces would imply the description of
these manifolds as
non-commutative spaces. Here we will give an alternative description in
terms of deformation quantization theory. We will be interested in compact
Poisson manifolds and in particular in symplectic and K\"ahler manifolds as
the spacetime of the underlying supersymmetric gauge theory.

Deformation quantization theory is the non-commutative
geometry for  Poisson
manifolds (for a recent review see \nref\weinstein{ A. Weinstein, ``Deformation
Quantization'', S\'eminaire Bourbaki 789, (1994) Asterisque {\bf 227} (1995)
389.}  \weinstein ). This theory deals with the quantization of
Poisson manifolds, that is, the
suitable deformation of Poisson structures on these manifolds
\ref\bayen{ F. Bayen, M. Flato, C. Fronsdal, A. Lichnerowicz and D. Sternheimer,
``Deformation Theory and Quantization I ,II'', Ann. Phys. {\bf 111} (1978) 61, 111.}.
The proof of the existence of quantization of {\it any} Poisson manifold has been
found recently in
\ref\kontsevich{ M. Kontsevich, ``Deformation Quantization of Poisson Manifolds, I'',
q-alg/9709040.} by using string theory techniques.
 Also deformation quantization can be explicitly carried over to {\it any}
finite dimensional curved symplectic manifolds with a symplectic connection
\ref\fedosov{B. V. Fedosov, ``A Simple Geometrical Construction of
Deformation Quantization'', J. Diff. Geom. {\bf 40} (1994) 213-238.}. These spaces
are called {\it Fedosov manifolds} and their differential geometry has been just recently
studied
\ref\gelfand{I. Gelfand, V. Retakh and M. Shubin, ``Fedosov Manifolds'',
dg-ga/9707024.}.

The purpose of this paper is to describe some conjectures regarding the geometry
of deformation quantization for matrix compactifications on the torus and on
curved manifolds using Fedosov geometry of deformation quantization theory.
We focus attention on the known BFSS model \bfss,
 though all
constructions can be applied straightforwardly to the IKKT model
\ref\Ikkt{Ishibashi, H. Kawai, Y.
Kitazawa and T. Tsuchiya, ``A Large-$N$
Reduced Model as Superstring'', Nucl. Phys. B {\bf 498} (1997) 467-491,
hep-th/9612115.}. The motivation of this generalization to curved manifolds is
the understanding of more general compactifications and the extension from
constant to non-constant background $B$-fields on the $T$-dual compact space $X$.
 This generalization was suggested in
\connes, where it is argued that the classical Lagrangian of the resulting deformed
gauge theory could be constructed following the lines of \ref\alanpaper{A. Connes,
Commun. Math. Phys. {\bf 182} (1996) 155, hep-th/9603053.}. In the present paper
we give an alternative construction of this Lagrangian in the context of
deformation quantization theory. This construction will be valid  for arbitrary
symplectic manifolds and indeed can be generalized to Poisson manifolds.

The organization of this paper is as follows: Sec. 2 is devoted to briefly
reviewing the derivation of super Yang-Mills theory on a non-commutative
$p$-dimensional dual torus \refs{\connes, \cheung,\kawano}. It is also shown
that a Weyl correspondence of Moyal quantization can be established  in this
framework and thus a deformed Lagrangian  can be obtained in the Moyal picture.
In Sec. 3 we show how the immediate  generalization to non-constant background
$B$-fields on the $T$-dual space involves the introduction of curved Poisson
manifolds as the spacetime
where the underlying gauge theory lives and it is exactly described
through the Fedosov geometry
of deformation quantization. In Sec. 3 we also study the gauge theory on a Fedosov
manifold and the corresponding Lagrangian is shown to describe
deformations of the gauge theory on a curved manifold parametrized by the
background $B$-field. The correspondence with the gauge theory
on the noncommutative torus is also given. Comments about the extension
to K\"ahler manifolds with symplectic connection and its application
to ALE and $K3$ spaces and Calabi-Yau threefolds are provided. Finally in Sec. 4 we give
our concluding remarks.

\vskip 2truecm


\newsec{Deformation Quantization of Matrix Compactifications on Tori}

In order to introduce some notation we will use in the paper we briefly review
the matrix theory action. It is given by a matrix quantum mechanics action with
SU$(N)$ gauge group and 16 supercharges  \ref\halpern{M. Claudson and M. Halpern,
Nucl. Phys. B {\bf 250} (1985) 689; R. Flume, Ann. Phys. {\bf 164} (1985) 189; M. Baake,
P. Reinicke and and V. Rittenberg, J. Math. Phys. {\bf 26} (1985) 1070.}

\eqn\action{I_M = \int d t L_M}
with
\eqn\lagra{
L_M = {1 \over g_{YM}^2}{\rm Tr}_N \Big(
 (D_t {\bf X}^I)(D_t {\bf X}^I) + {1\over 2} [{\bf X}^I , {\bf X}^J]
[{\bf X}^I , {\bf X}^J]+ i{\bf \Theta}^{\alpha}D_t {\bf \Theta}^{\alpha} -
{\bf \Theta}^{\alpha} \Gamma^I_{\alpha \beta } [{\bf X}_I ,
{\bf \Theta}^{\beta}] \Big)}
where ${\bf X}^I$ and ${\bf \Theta}^{\alpha}$ are ($N \times N$ matrix-valued)
9 bosonic and 16 spinor
coordinates of 0-brane partons ($I = 1, \cdots, 9$ are SO$(9)$ indices with
metric $\delta_{IJ}$ and $\alpha = 1, \cdots,
16$ are SO$(9)$ spinor indices), ${\rm Tr}_N$ is the trace in the fundamental
representation of SU$(N)$. The Majorana spinor conventions are such that the
$\Gamma_I$'s
 are real
and symmetric and obey $\{\Gamma^I,\Gamma^J\}= 2 \delta^{IJ}$, $\Gamma^1 \dots
\Gamma^9 = + {\bf 1}$ and $\Gamma^{IJ} = {1 \over 2} [\Gamma^I, \Gamma^J]$.
As $N$ goes to infinity  Lagrangian \lagra\ describes membranes $\Sigma$
extending along $t,x^{11}$ directions \bfss.

The
gauge covariant
 derivative reads $D_t {\bf X}^I \equiv \partial_t
{\bf X}^I - i [A_t,{\bf X}^I]$ and
$D_t {\bf \Theta}^{\alpha} \equiv \partial_t
{\bf \Theta}^{\alpha} - i [A_t,{\bf \Theta}^{\alpha}].$ These definitions
ensure invariance under area-preserving diffeomorphism group SDiff$(\Sigma)$.

\subsec{Non-commutative Torus and D0-branes with Background $B$-field}

Recently it has been shown that matrix theory compactifications on
${\bf T}^2$ with a {\it constant} background three-form tensor field $C_{-ij}$,
can be described in terms of a $(2+1)$-dimensional super Yang-Mills theory on a
non-commutative dual torus $\tilde{\bf T}_{\theta}^2$ \connes\ . More recently
the analogous correspondence in the context of $D0$-branes of type IIA
superstrings compactified on
${\bf T}^p$ has been done \refs{\cheung,\kawano}. In these works
the corresponding $(p+1)$-dimensional super Yang-Mills action on the
noncommutative dual
$p$-torus, $\tilde{\bf T}_{\theta}^p$ was found. This was done by placing a set of $N$
$D0$-branes on the space ${\bf T}^p$,
which is a flat space ${\bf R}^p$ with ${\bf Z}^p$ periodicity.

It is known that
the ${\bf X}^I$ and ${\bf \Theta}^{\alpha}$ must satisfy the known constraints
of $D$-branes on tori \refs{\taylor,\ganor}

\eqn\condu{{\bf X}^A_{m,n} = {\bf X}^A_{m-n,0}, \ \ \ \ \ \ \  A= p+1, \dots ,9 }

\eqn\condd{{\bf X}^i_{m,n} = {\bf X}^i_{m-n,0} + 2 \pi \ell_s E^i_b n^b
\delta_{m,n},
\ \ \ \  i=1, \dots ,p }

\eqn\condt{{\bf \Theta}^{\alpha}_{m,n} = {\bf \Theta}^{\alpha}_{m-n,0}}
where $n,m$ are integers and label the positions of the $D0$-branes or
the initial and final points of the corresponding string on the lattice
${\bf Z}^p$,
$\ell_s$ is the string scale, $E^i_a$ is the vielbein associated to the metric
on the flat torus $G_{ab} = E^i_a E^j_b \delta_{ij}$.

The $T$-dual picture enables the existence
of $N \times N$ hermitian matrix-valued functions through the Fourier transform
\refs{\taylor,\ganor}. Thus the correspondence between matrices  ${\bf X}^I(t)$
and matrix-valued functions on the torus, ${X}^I(t,\tilde{\sigma})$ is given by
\nref\hoppe{J. Hoppe, M.I.T. Ph.D. Thesis (1982); Phys. Lett. B {\bf 215} (1988) 706.}
\nref\ffz{D.B. Fairlie, P.. Fletcher and C.K. Zachos, J. Math. Phys. 31
(1990) 1088.}

\eqn\weyl{ \sigma_N^{-1}: Mat_N \to C^{\infty} (\tilde{\bf T}^p)\otimes Mat_N}
where $Mat_N$ is the set of $N \times N$ non-singular matrices representing the
Lie algebra su$(N)$ and
$C^{\infty}(\tilde{\bf T}^p)$ is the set of smooth functions on the dual torus
$\tilde{\bf T}^p$. The map
$\sigma_N^{-1}$ is given by the Fourier transformation

\eqn\map{ { X}^I(t, \tilde{\sigma}) = \sigma_N^{-1}({\bf X}^I) := \sum_{n \in {\bf Z}^p}
{\bf X}^I_{n,0} (t,N) exp( in^j \tilde{\sigma}_j)}
where
\eqn\fourier{ {\bf X}^I_{n,0}(N,t) = {1 \over (2 \pi)^p} \int_{\tilde{\bf T}^p}
 d^p \tilde{\sigma}
X^I(t, \tilde{\sigma}) exp \big( -i n^j\tilde{\sigma}_j  \big) }
where $\tilde{\sigma} = (\tilde{\sigma}_1, \dots, \tilde{\sigma}_p)$ and
$\tilde{\sigma}_j$ $(j =1, \dots
,p)$ are the local coordinates on $\tilde{\bf T}^p$. For the fermionic counterpart there
exist similar expressions. The inverse transformation $\sigma_N$ is well defined
as it is given by the inverse Fourier transformation. Constraints (2.3-2.5) can
be solved in the $T$-dual picture replacing the matrices ${\bf X}^i(t)$ by the
matrix-valued functions on the dual tori ${X}^i(t,\tilde{\sigma}) = i
\partial^i + { A}^i(t, \tilde{\sigma})$ and thus the correspondence between
matrix compactifications on the tori and gauge theories with 16 supercharges
can be finally established \refs{\taylor,\ganor}.

\vskip 1truecm
\subsec{ Moyal Deformation in Toroidal Compactifications}

Now we are going to repeat the same correspondence in the presence of a
non-vanishing background $B$-field.
It was shown in \refs{\cheung,\kawano}, that a non-vanishing constant background
$B$-field introduces the factor

\eqn\evolution{ U = exp \bigg( - i {\pi} B_{ij} n^i m^j \bigg)}
for each interaction term of three strings on the lattice. This is because
the closed two-form $B_{ij}$ couples to the worldsheet under the
interaction $\int_{WS} B$. This coupling depends only on the homotopy type of the
worldsheet embedding and is of the form $ exp({i \over 2}B)
\phi^{(3)}_{ik}\phi^{(2)}_{kj}
\phi^{(1)}_{ji}$ where $\phi^{I} = (\phi^i, \phi^A)$ denotes a generic field of the
theory.
Generalization to interaction of $k$ strings on the lattice leads to the term
$Tr\big( \phi^{(k)} * \phi^{(k-1)} * \dots *\phi^{(2)}*\phi^{(1)} \big)$
where the $*$-product is given by \cheung

\eqn\diferente{ \big( \phi^{(2)} * \phi^{(1)}\big)_{a_3b_3,a_1b_1} =
\sum_{a_2,b_2} \phi^{(2)}_{a_3b_3,a_2b_2}
exp \bigg( {1\over 2} B \left|\matrix{
a_3 - a_2& a_2 - a_1 \cr
b_3 - b_2& b_2 - b_1 \cr
}\right| \biggl) \phi^{(1)}_{a_2b_2,a_1b_1}. }
The subindices of the fields $\phi$ denote the positions of the $D0$-branes
on the lattice. Solving
the constraints (2.3-2.5), the $*$-product \diferente\ in the dual basis
$\{\tilde{\sigma}\}$ looks
like \refs{\connes,\cheung,\kawano}

$$\phi^I(\tilde{\sigma}) * \phi^J (\tilde{\sigma})\equiv \phi^I(\tilde{\sigma}){\rm exp}
\bigg( + i \pi \ell_s^2  B^{ij}
{{\buildrel{\leftarrow} \over
\partial} \over \partial \tilde{\sigma}^i}{{\buildrel{\rightarrow} \over
\partial} \over \partial
\tilde{\sigma}^j}\bigg)\phi^J(\tilde{\sigma})$$

\eqn\moyal{= \sum^{\infty}_{k=0} ( i \pi \ell_s^2)^k {1 \over k!} B^{i_1j_1}
\ldots B^{i_kj_k} {\partial^k \phi^I(\tilde{\sigma})\over \partial
\tilde{\sigma}^{i_1}\ldots
\partial \tilde{\sigma}^{i_k}} {\partial ^k \phi^J(\tilde{\sigma}) \over \partial
 \tilde{\sigma}^{j_1}
 \ldots \partial
\tilde{\sigma}^{j_k}}}
for all $\phi^J(\tilde{\sigma})\in C^{\infty}(\tilde{\bf T}^p) \otimes Mat_N$.
This is precisely the associative and noncommutative Moyal product of
 deformation quantization theory \bayen. Thus
the effect of turning on the tensor $B$-field on the lattice  deform the
usual product of functions on $\tilde{\bf T}^p$ and turning it into the
Moyal product, with a
deformation parameter $\zeta_B$ given by the constant components of the
$B$ tensor field and the constant $\pi \ell^2_s$.

From now on we leave aside the $Mat_N$ sector the right hand side of \weyl. We
will consider only the $C^{\infty}(\tilde{\bf T}^p)$ part. At the end we will
consider the matrix-valued structure taking the trace Tr$_N$ in all relevant
equations. Then all fields in the Yang-Mills theory are matrix-valued
functions on the $p$-torus and they are written as

\eqn\ara{ { \phi}^I(t, \tilde{\sigma}) = \sum_{n \in {\bf Z}^p}
{\bf \Phi}^I_{\bf n} (t,N) L_{\bf n}.}
where $L_{\bf n} \equiv exp( in^j \tilde{\sigma}_j).$

It is well known that the basis of
the Lie algebra su$(N)$ can be seen as a two-indices infinite algebra for
$p=2$  \refs{\hoppe,\ffz}. The elements of this basis are denoted by $L_{\bf m}$,
$L_{\bf n}$,...,
etc., ${\bf m} = (m_1,m_2)$, ${\bf n} = (n_1,n_2)$,..., etc., and ${\bf
m}, \ {\bf n},...  \in I_N \subset {\bf Z} \times {\bf Z} - \{(0,0) \} \ {\rm
mod} \ N{\bf q},$ where ${\bf q}$ is any element of ${\bf Z} \times
{\bf Z}$. The basic vectors $L_{\bf m}, \ {\bf m}\in I_N,$ are the $N
\times N$ matrices satisfying the following commutation relations

\eqn\algebra{ [L_{\bf m}, L_{\bf n}] = {N \over \pi} {\rm sin} \big( {\pi \over N}
{\bf m} \times {\bf n} \big) L_{{\bf m} + {\bf n} \ \ {\rm mod} \ \  N{\bf
q}},}
where ${\bf m} \times {\bf n} := m_1n_2 - m_2 n_1$.  For the 2-torus $\tilde
{\bf T}^2$ the generators read $L_{\bf n} \equiv exp \big(i(n_1 \tilde{\sigma}_1
 + n_2 \tilde{\sigma}_2) \big).$ Large $N$ limit ($N \to \infty$) of algebra
 \algebra\ gives the area-preserving diffeomorphism algebra
sdiff$(\tilde{\bf T}^2).$

The correspondence \weyl\ can be seen as the composition of two mappings.
The first one is a Lie algebra representation of su$(N)$ (for {\it finite} $N$)
into a
Lie algebra  $\hat{\cal G}$ of self-adjoint operators acting on the Hilbert space
$L^2({\bf R}),$ given by

\eqn\une{{\bf \Psi}: {\rm su}(N) \to \hat{\cal G},
 \ \ \ \ \  {\bf \Phi} \mapsto {\bf \Psi}({\bf \Phi}^J) := \hat{\bf \Phi}^J.}
This map can be constructed explicitely for $N=2$ \nref\maciej{
J.F. Pleba\'nski, M. Przanowski and H. Garc\'{\i}a-Comp\'ean, Mod. Phys. Lett.
A, {\bf 11} (1996) 663, hep-th/9509092.} \maciej.
The second mapping is a genuine
{\it Weyl correspondence} ${\cal W}^{-1}$ which establishes a one to one
correspondence between the algebra ${\cal B}$ of self-adjoint linear operators
acting on
$L^2({\bf R})$ and the space of real smooth functions
$C^{\infty}(\tilde{\bf T}^2)$ where $\tilde{\bf T}^2$ is seen as the classical
phase-space. This
correspondence $ {\cal W}^{-1}: {\cal B} \to C^{\infty}(\tilde{\bf T}^2),$
is given  by

\eqn\corresp{\phi^J(t,\tilde{\sigma}^1,\tilde{\sigma}^2;\zeta_B) \equiv
{\cal W}^{-1}(\hat{\bf \Phi}^J) :=
\int_{- \infty}^{\infty} <\tilde{\sigma}^1 - {\xi \over 2}|
\hat{\bf \Phi}^J(t)|\tilde{\sigma}^1 + {\xi \over
2}> {\rm exp}\big( {i \over \zeta_B} \xi \tilde{\sigma}^2 \big) d\xi,}
for all $\hat{\bf \Phi}^J \in {\cal B}$ and ${\phi}^J \in C^{\infty}
(\tilde{\bf T}^2).$
Thus from the identification of ${\cal B}$ with $\hat{\cal G}$, it follows that
the correspondence $\sigma_N^{-1}$ is equal to the map composition
$\sigma^{-1}_N = {\cal W}^{-1} \circ {\bf \Psi}$ for finite $N$.

With the mapping ${\bf \Psi}$ and the Weyl correspondence \corresp\ it
is easy to check that
\eqn\homo{ \sigma_N^{-1} \big( {\bf \Phi}^I{\bf \Phi}^J \big) = \phi^I * \phi^J}
where $\sigma_N^{-1} ({\bf \Phi}^I) := \phi^I$ and ${\bf \Phi}^I \in Mat_N$. One
can see that $\sigma_N^{-1}$ is actually a Lie algebra isomorphism

\eqn\mapeo
{\sigma_N^{-1} : (Mat_N, [\cdot,\cdot]) \to ({\cal M}_B, \{\cdot,\cdot\}_B)}
where ${\cal M}_B$ is a family of associative algebras on a fixed complex
vector space denoted by ${\cal M} = C^{\infty}(\tilde{\bf T}^p)$. This isomorphism
leads to the definition of the Moyal bracket

\eqn\homo
{ \sigma_N^{-1} \big({1 \over i \zeta_B} [{\bf \Phi}^I,{\bf \Phi}^J] \big)
= {1 \over i \zeta_B}\big(\phi^I(\tilde{\sigma})
* \phi^J(\tilde{\sigma}) - \phi^J(\tilde{\sigma}) * \phi^I(\tilde{\sigma})
\big) \equiv
\{ \phi^I(\tilde{\sigma}), {\phi}^J(\tilde{\sigma}) \}_B}
where $\{\cdot,\cdot\}_B$ is the Moyal bracket and $[\cdot,\cdot]$ is the usual
commutator of matrices.

The algebra of quantum
observables

\eqn\alg{({\cal M}_B, \{\cdot,\cdot\}_B)}
can be defined more precisely by introducing an
associative $*$-product operation
on the vector space ${\cal M}$ of functions $\phi^I(\tilde{\sigma},\zeta_B) =
\sum_{k=0}^{\infty} \zeta_B^k \phi^I_k(\tilde{\sigma})$ and
 $\phi^J(\tilde{\sigma},\zeta_B)
 = \sum_{k=0}^{\infty}
\zeta_B^k \phi^J_k(\tilde{\sigma})$, with $\phi^I_k(\tilde{\sigma}),
\phi^J_k(\tilde{\sigma})  \in {\cal M}$.
The $*$-product is defined by $\phi^I*\phi^J=\phi^K =
\sum_{k=0}^{\infty}\phi^K_k(\tilde{\sigma})$ for all $\phi^I,\phi^J,\phi^K \in
{\cal M}$ satisfying the
properties $(i)$-. $\phi^K_k$ are polynomials in $\phi^I_k$ and $\phi^J_k$ and their
derivatives. $(ii)$-. $\phi^K_0(\tilde{\sigma}) = \phi^I_0(\tilde{\sigma})
\phi^J_0(\tilde{\sigma})$.  $(iii)$-. $\{\phi^I,\phi^J\}_B \equiv
{1 \over i \zeta_B}(\phi^I*\phi^J -
\phi^J * \phi^I) = \{\phi^I_0,\phi^J_0\}_P + ...,$ where
$\{\cdot,\cdot\}_P$ stands
for the Poisson bracket and the dots mean the terms of higher orders.  In
order to be
more precise
${\cal M}$ is a linear space whose elements are of the form $ \phi^I=\phi^I(\tilde{\sigma},
 \zeta_B)=
\sum_{k=0}^{\infty} \zeta_B^k \phi^I_k(\tilde{\sigma}),$ where $\phi^I_k(\tilde{\sigma})
 \in C^{\infty}(\tilde{\bf T}^p)$.

We have seen that a background $B$-field is projected over the dual torus where
it is defined as
an antisymmetric tensor field. It can also be seen as a non-degenerate
closed two-form
on the torus $\tilde{B}=B_{ij} d\tilde{\sigma}^i \wedge d\tilde{\sigma}^j$.
For $p$ even, $B_{ij}$
can be identified with the symplectic two-form on the dual torus, {\it i.e.}
$\omega_{ij}=B_{ij}$. For instance for $p=2$ symplectic form on $\tilde{\bf T}^2$ is
$\epsilon_{ij}$, while the $B$-field is given by $B_{ij} = B \epsilon_{ij}$ with $B$
a constant value. The deformation parameter $\zeta_B$ is identified with
$\pi \ell^2_s B$ for fixed $\ell_s$. Jacobi identity follows from the closeness
condition of $\tilde{B}$, {\it i.e.} $d\tilde{B} = 0$. Poisson structure
on $\tilde{\bf T}^2$ is
given by $\{\phi^I,\phi^J\}_P
={B^{ij}\over B} \partial_i \phi^I \partial_j \phi^J.$ For $p>2$ the symplectic structure
is identified with $B_{ij}$. For any valued of $p$ (even or odd) $B_{ij}$
determines a Poisson structure on the underlying compact space.

\vskip 1truecm
\subsec{ Gauge Theory on the Non-commutative Torus}

Finally the $B$-deformed Lagrangian of the gauge theory can
be written as
$$
L_M = M_P\int_{\tilde{\bf T}^p} d^p \tilde{\sigma} {\rm Tr}_N
 \bigg( - {1 \over 4 T^2} F^{\mu \nu} *F_{\mu \nu } -
{1 \over 2 } (D_{\mu} X^A) * (D_{\mu}{X}^A) + {T^2 \over 4}
\{{X}^A , {X}^B\}_B * \{{X}^A , {X}^B\}_B $$

\eqn\lagrangiano
{ - {i \over 2}{\Theta}^{\alpha} * \Gamma^{\mu}_{\alpha \beta} D_\mu
{\Theta}^{\beta} + {T\over 2}
{\Theta}^{\alpha} *\Gamma^A_{\alpha \beta }\{{X}_A,{\Theta}^{\beta}\}_B \bigg)}
where $\mu, \nu = 0, \dots, p$. Fields are now ${\cal M}$-valued fields on the
$(p+1)$-dimensional space time ${\bf R} \times \tilde{\bf T}^d$. Gauge fields on the
dual torus are defined as  $ A^i(t,\tilde{\sigma}) := {1 \over 2 \pi \ell_s^2} X^i
(t,\tilde{\sigma})$. $A_{\mu} =A_{\mu}(t,\tilde{\sigma})$  are composed by the
$A^0$ and $A^i$ fields where we redefine $\tilde{\sigma}^{\mu} \equiv (\tilde{\sigma}^0,
\tilde{\sigma})$ with $t \equiv \tilde{\sigma}^0.$

The field strength of these fields is

\eqn\strength{F_{\mu \nu}(\tilde{\sigma}) = {\partial \over \partial \tilde{\sigma}^{\mu}}
A_{\nu}(\tilde{\sigma}) - {\partial \over \partial \tilde{\sigma}^{\nu}}
A_{\mu}(\tilde{\sigma}) +  \{ A_{\mu}(\tilde{\sigma}),A_{\nu}(\tilde{\sigma})\}_B.}
The scalar fields are $X^A(\tilde{\sigma})$ and their coupling to the gauge fields
is given through the covariant derivative

\eqn\covariant{D_{\mu}
X^A(\tilde{\sigma}) = {\partial \over \partial \tilde{\sigma}^{\mu}}
X^A(\tilde{\sigma}) +   \{ A_{\mu}(\tilde{\sigma}),X^A(\tilde{\sigma})\}_B.}
Similar expressions hold for the Majorana spinors $\Theta^{\alpha}(\tilde{\sigma})$
\eqn\fermion{D_{\mu}
\Theta^{\alpha}(\tilde{\sigma}) = {\partial \over \partial \tilde{\sigma}^{\mu}}
\Theta^{\alpha}(\tilde{\sigma}) +   \{ A_{\mu}(\tilde{\sigma}),\Theta^{\alpha}
(\tilde{\sigma})\}_B.}

We have obtained the $(p+1)$-dimensional Yang-Mills theory with maximal
supersymmetry
on the noncommutative space ${\bf R} \times\tilde{\bf T}^p_B$.
Its bosonic part was
worked out some years ago in
\nref\floratos{ E.G. Floratos, J. Iliopoulos and G. Tiktopoulos, Phys. Lett.
{\bf B217} (1989) 285. } \nref\bars{ I. Bars, ``Strings From Reduced Large-$N$
Gauge Theory Via
Area-Preserving Diffeomorphisms'', Preprint USC-90/HEP-12,
IASSNS-HEP-90/21, (1990); ``Area Preserving Diffeomorphisms as a Bridge
Between Strings and Large-$N$ QCD'', Proceedings, Strings 1990 pp.
202-209.}
\nref\caswzw{C. Castro, J. Math. Phys. {\bf 35}
(1994) 920.}
\nref\jerzy{H. Garc\'{\i}a-Compe\'an and  J.F. Pleba\'nski, Phys. Lett. A {\bf 234}
(1997) 5; H. Garc\'{\i}a-Compe\'an, J.F. Pleba\'nski and N. Quiroz-P\'erez,
``On the Reduced SU$(N)$ Gauge Theory in the WWM-Formalism'', to appear in
Int. J. Mod. Phys. A, hep-th/9610248.}
\refs{\ffz,\floratos,\bars,\caswzw,\jerzy}.

One of the advantages of deformation quantization is that it permits a unified view of
the mentioned correspondence between matrix compactifications and gauge theories. In the
case considered in this paper, the non-commutative gauge theory Lagrangian
\lagrangiano\ can be
derived straightforwardly from the Moyal $B$-deformation of the matrix theory Lagrangian
\lagra. The latter Lagrangian was proposed in Ref.
\ref\fairlie{D.B. Fairlie, ``Moyal Bracket in $M$-theory'', Mod. Phys. Lett.
A {\bf 13} (1998) 263, hep-th/9707190.}. The
derivation involves the solution of the functional version of constraints (2.3-2.5).
Lagrangian \lagrangiano\ is not expected to possess global anomalies (of the
type of
\ref\baadhio{R.A. Baadhio, ``Global Anomalies Induced by Deformed
Quantization'', Preprint UCB-PTH-94/20; {\it Quantum Topology and Global
Anomalies} World Scientific, (1996).})
coming from the
degeneracy of the Poisson bracket if $B_{ij}$ is non-degenerate.

Noncommutative gauge field theories \lagrangiano\ are non-local field theories of the type
described in \berkooz. These theories have no renormalization group fixed point in
the UV. However the theory  is still well defined in the continuum. In \berkooz\ a manner
to study the renormalization group by taking a limit in which the
size of the torus goes to infinity, while the size of the non-locality scale
is keeping fixed, was
proposed . Six-dimensional (0,2) field theories in the DLCQ description
do admit a generalization of
this type. At the end of Sec. 4 we will return to this point.

\vskip 2truecm

\newsec{ Geometry of Deformation Quantization in Matrix Compactifications}

The purpose of this section is to formulate the gauge theory underlying matrix
compactification on tori,
in terms of noncommutative geometry of the dual torus $\tilde{\bf T}^p$ with a
symplectic structure given by a $B$ field.
We will use for this the
deformation quantization theory given by  de Wilde and Le Compte
\nref\wilde{M. de Wilde and P.B.A. Le Compte, Lett. Math. Phys. {\bf 7}
(1983) 487-496.} \wilde\  and Fedosov \fedosov\ . We will find in
deformation quantization
theory a natural framework to generalize to non-constant background
$B$-field on
more general curved symplectic $T$-dual manifolds $X$.
That means, $B$-fields depending on the point $x$
in
$X$. For each such a point we can associate a Weyl algebra defined on the
tangent bundle $TX$  to $X$, to the Moyal algebra \alg.
This gives precisely a bundle structure known in deformation
quantization
theory as  {\it Weyl algebra bundle} \refs{\wilde,\fedosov}. Gauge theory on the
$T$-dual curved space $X$ will be found and we will rederive the toroidal
result of Sec. 2 as an special limit when Riemannian curvature of $X$ vanishes.
Finally some comments about the application of deformation quantization formalism
to K\"ahler manifolds is also given.

\vskip 1truecm
\subsec{Deformation Quantization}
Before we start with the application of deformation quantization theory to matrix
compactifications it is convenient to recall some of its terminology
(for a nice review see
\weinstein\ ). The aim of deformation quantization program is the description of
deformation of a Poisson algebra ${\cal A}$, ${\cal A}_{\hbar},$ associated to
some Poisson manifold $X$
through a family of deformed product of functions $*_{\hbar}$. Here $\hbar$ is the
deformation parameter. Equivalent description can be done by means of a sequence of bilinear
mappings $M_k: {\cal A} \times {\cal A} \to {\cal A},$   $k =0,1,\dots$, with

\eqn\producto{ a *_{\hbar} b = \sum_k \hbar^k M_k(a,b)}
where $a,b \in {\cal A}_{\hbar}.$

The problem of formal deformation quantization is to classify such a families up to
equivalence of the $*_{\hbar}$-product.

The most part of the realization of this program has been done in the context of
algebraic
structures of quantum mechanics \bayen. This can be implemented with the obvious
identifications
${\cal A} = C^{\infty}(X)$, $X$ the classical phase space and the realization
of  $M_0(a,b)$ as the usual
product of functions $ab$ and linear combinations of $M_1(a,b)$ as the Poisson bracket.

One interesting example of Poisson manifold is a symplectic manifold $(X,\omega).$
 The symplectic structure is defined in terms of the
Poisson structure on $X$. {\it Locally} the symplectic form can always be  written
as $\omega = \sum_i dq_i \wedge dp_i$, where $\{x^i\}$ $i=1,\dots,2N$ with $x^i=p^i,$
$i\leq N$ and $x^i = q^{i-N},$ $i>N$, are the local coordinates on
$X$.

Globally the symplectic form is defined by $\omega : TX \to T^*X$ with inverse
$\omega^{-1}:T^* X \to TX$.  Here $TX$
and $T^*X$ are the respective tangent and cotangent bundles to
$X$. While the hamiltonian (or volume preserving) vector fields are
$V_{H_a} = \omega^{-1}(dH_a)$ satisfying the sdiff$(X)$
algebra $[V_{H_a},V_{H_b}] = V_{\{H_a,H_b\}_P} \ \ \ ({\rm
for \ all } \ a\not = b),$
where $\{\cdot,\cdot \}_P$ stands
for the Poisson bracket with respect to $\omega$. Locally it can be
written as $\{H_a,H_b\}_P = \omega^{-1}(dH_a,dH_b) = \omega^{ij} \partial_i H_a
\partial_j H_b, $
where $\partial_i \equiv {\partial \over \partial x^i}$ and
$H_i= H_i(x).$
The generators of sdiff$(X)$ are the hamiltonian vector fields $V_{H_a}$
associated to the hamiltonian functions $H_a$ and they form a Lie algebra
whose Lie group is the volume-preserving diffeomorphism group SDiff$(X)$ of
phase space $X$.

Moyal product for symplectic manifolds with Poisson structure given by \producto\ is
characterized by the sequence

\eqn\coef{ M_k(a,b) (x) = {1 \over k!} \bigg({i \over 2} \sum_{i,j} a(y)
\big(\omega_{ij} { \buildrel{\leftarrow} \over {\partial}   \over
\partial y_i} {\buildrel{\rightarrow} \over {\partial}
 \over \partial z_j}\big)^k b(z) |_{y=z=x} \bigg)}
where $\omega_{ij} = \omega(x_i,x_j)$. This sequence involves only differential
operators which define $*$-product only locally. The $*$-product is thus in general
not globally defined on $X$. Global $*$-product exists always for {\it any} finite
dimensional symplectic manifold \wilde. Among other proofs of the existence theorem,
that of \ref\omy{
H. Omory, Y. Maeda and A. Yoshioki, ``Weyl Manifolds and Deformation
Quantization'', Advances in Math. {\bf 85} (1991) 224.} involves the construction
of a different constant Poisson structure for each tangent space on $X$. The
tangent bundle $TX$ becomes a Poisson manifold with fiberwise Poisson bracket and
with fiberwise quantization. The quantization of $X$ is given by the induced
multiplication on $C^{\infty}(X)[[\hbar]]$ from the multiplication on
$C^{\infty}(TX)[[\hbar]]$ or Weyl structure  on $X$. A existence proof given
in \wilde\ uses methods of C$\check{e}$ch cohomology of $X$. A classification of
$*$-products in terms of C$\check{e}$ch cohomology using gerbes theory was done
in
\ref\deligne{ P. Deligne, ``D\'eformations de l'Alg\`ebre des Fonctions d'une
Vari\'et\'e Symplectique: Comparaison entre Fedosov et De Wilde, Lecomte'',
Selecta Mathematica {\bf 1} (1995) 667.}. A treatment of deformation
quantization parallel to non-commutative theory methods is given in
\ref\gozzi{ E. Gozzi and M. Reuter, ``Quantum-Deformed Geometry on
Phase-Space'', Int. J. Mod. Phys. A {\bf 11} (1996) 1253-1278, hep-th/9510011.
M. Reuter, ``Non-Commutative Geometry on Quantum Phase-Space'',
Int. J. Mod. Phys. A {\bf 11} (1996) 1253; M. Reuter, ``Noncommutative Geometry,
 Multiscalars and the Symbol
Map'', Nucl. Phys. B Proc. Suppl. {\bf 49} (1996) 333-337.}. Recently Kontsevich
gave a proof of existence of the global $*$-product for every
Poisson manifold using techniques of string theory and topological field
theories in two-dimensions \kontsevich.
Finally a very clear
perspective of the problems arising in the mathematical theory of
deformation quantization was given
recently by Rieffel in \ref\rieffel{ M.A. Rieffel, ``Questions on Quantization'',
quant-ph/0712009.}.

\vskip 1truecm

\subsec{Fedosov's Geometry of Deformation Quantization}

In this subsection we discuss the elements of Fedosov's geometry of deformation
quantization theory. Some review of this subject can be found in
\nref\emmrich{ C. Emmrich and A. Weinstein, ``The Differential Geometry of Fedosov's
Quantization'', hep-th/9311094.}
\refs{\weinstein,\emmrich,\deligne}. We will follow the discussion for the torus
but we will specify whenever general formulas for the space $X$
be required.
\bigskip

\noindent
{\it Weyl algebra bundle}

We consider $X$ to be the dual torus $\tilde{\bf T}^p$ (with $p$ even)
and symplectic structure given by the
tensor $B_{ij}$. $\tilde{\bf T}^p$ has a natural Riemannian
structure given by the flat metric $\eta_{ij}.$

The formal Weyl
algebra ${\cal W}_{\tilde{\sigma}}$ associated with the tangent space
$T_{\tilde{\sigma}} \tilde{\bf T}^p$ at the point
$\tilde{\sigma} \in
\tilde{\bf T}^p$  is the
{\it associative} algebra over {\bf C}
with a unit. An element of ${\cal W}_{\tilde{\sigma}}$ can be expressed by
$\tilde{\phi}^I(y) = \sum_{2k +l \geq 0} \zeta_B^k \tilde{\phi}^I_{k, i_1 \ldots i_l}
 y^{i_1} \ldots
y^{i_l},$ where $\zeta_B$ is the deformation parameter,
$y = (y^1, \ldots ,y^{2n})$ is a tangent vector and the coefficients
$\phi^I_{k,i_1 \ldots
i_l}$ constitute the symmetric covariant tensor of degree $l$ at
$\tilde{\sigma} \in
 \tilde{\bf T}^p$. The product on ${\cal W}_{\tilde{\sigma}},$ which
 determines the associative algebra structure is defined by

$$ \tilde{\phi}^I\circ \tilde{\phi}^J\equiv \tilde{\phi}^I (y,\zeta_B) {\rm exp}
 \bigg( +i {\zeta_B \over 2} B^{ij}
{ \buildrel{\leftarrow}\over {\partial}\over \partial y^i}
{\buildrel{\rightarrow}\over {\partial}\over \partial z^j}
\bigg)  \tilde{\phi}^I
 (z, \zeta_B) |_{z=y}
$$

\eqn\fed{= \sum^{\infty}_{k=0} ( +i {\zeta_B \over 2})^k {1 \over k!} B^{i_1
j_1} \ldots B^{i_kj_k} {\partial^k \tilde{\phi}^I\over \partial y^{i_1} \ldots
\partial y^{i_k}} {\partial ^k \tilde{\phi}^J \over \partial y^{j_1} \ldots \partial
y^{j_k}},}
for all $\tilde{\phi}^I, \tilde{\phi}^J \in {\cal W}_{\tilde{\sigma}}$.
Here $B^{ij}$ are the components of
the tensor inverse to $B_{ij}$ at $\tilde{\sigma}$. Of course the product
``$\circ$'' is independent of the basis. Thus one can define an
 algebra bundle structure taking the disjoint
union of Weyl algebras for all points $\tilde{\sigma} \in \tilde{\bf T}^p \
 i.e. \ \tilde
{\cal W} = {\parallel \over \tilde{\sigma} \in \tilde{\bf T}^p}
{\cal W}_{\tilde{\sigma}}$. $\tilde{\cal W}$ is the
total space and the fiber is isomorphic to a Weyl algebra
${\cal W}_{\tilde{\sigma}}$.
Thus we have the Weyl algebra bundle structure

\eqn\bundle{\tilde{\cal W}
\buildrel{\pi}\over{\to} \tilde{\bf T}^p, \ \ \ \ \ {\cal W}_{\tilde{\sigma}}
\cong \pi^{-1}(\{\tilde{\sigma}\}),}
where $\pi$ is the canonical projection.

Let ${\cal E}( \tilde {\cal W})$ be the set of sections of $\tilde {\cal
W}$ which also has a Weyl algebra structure with unit. Denote by $\tilde{\phi}^I
(\tilde{\sigma},
y,\zeta_B)$ an element of ${\cal E} (\tilde {\cal W})$, it can be written as
follows

\eqn\forma{ \tilde{\phi}^I (\tilde{\sigma}, y, \zeta_B) =
 \sum_{2k + l \geq 0} \zeta_B^k \
\tilde{\phi}^I_{k, i_1 \ldots i_l}
(\tilde{\sigma}) y^{i_1} \ldots y^{i_l},}
where $y = (y^1,...,y^{2n}) \in T_{\tilde{\sigma}} \tilde{\bf T}^p$ is a tangent vector,
$\tilde{\phi}^I_{k,i_1 \ldots i_l}$ are smooth functions on $\tilde{\bf T}^p$ and
$\tilde{\sigma} \in \tilde{\bf T}^p$.

Fedosov's deformation quantization theory also permits the definition of Weyl algebra-valued
differential forms  on $\tilde{\bf T}^p$. Such a $p$-form is defined by

\eqn\field{\tilde{\phi}^I =\sum_{2k + p \geq 0}\zeta_B^k \tilde{\phi}^I_{k, j_1...j_p}
(\tilde{\sigma},y)
d\tilde{\sigma}^{j_1}\wedge ... \wedge
d\tilde{\sigma}^{j_p}}
where $\tilde{\phi}^I_{k, j_1...j_p}(\tilde{\sigma},y) =
\tilde{\phi}^I_{k, i_1...i_l,j_1...j_p}(\tilde
{\sigma}) y^{i_1}...y^{i_l}$.

The set of differential forms constitutes a
{\it Grassmann - Cartan} algebra ${\cal C}= {\cal E} \big( \tilde W
\otimes \Lambda \big)= \bigoplus_{q=0}^{2n} {\cal E}\big( \tilde W \otimes
\Lambda^q\big)$. In this space the multiplication
$\tilde{\phi} \buildrel{\circ}\over{\wedge} \tilde{\phi}$ is defined by

\eqn\product{\tilde{\phi}^I \buildrel{\circ}\over{\wedge} \tilde{\phi}^J =
\tilde{\phi}^I_{[j_1\ldots j_p} \circ \tilde{\phi}^J_{l_1
\ldots l_q]} d\tilde{\sigma}^{j_1} \wedge \ldots \wedge d\tilde{\sigma}^{j_p}
\wedge d\tilde{\sigma}^{l_1} \wedge
\ldots \wedge d\tilde{\sigma}^{l_q},}
for all $\tilde{\phi}^I = \sum_k \zeta_B^k \tilde{\phi}^I_{k,j_1 \ldots j_p}
(\tilde{\sigma},y) d\tilde{\sigma}^{j_1}\wedge
\ldots \wedge d\tilde{\sigma}^{j_p} \in {\cal E}\big(\tilde{\cal W} \otimes
\Lambda^p\big)$ and $\tilde{\phi}^J = \sum_k \zeta_B^k \tilde{\phi}^J_{k, l_1 \ldots l_q}
(\tilde{\sigma},y)$
$d\tilde{\sigma}^{l_1}\wedge \ldots \wedge d\tilde{\sigma}^{l_q} \in {\cal E}
\big ( \tilde{\cal W}
\otimes \Lambda^q \big)$. The $\circ$-wedge product is defined
by the usual wedge product on $\tilde{\bf T}^p$ and the  $\circ$-product in the
Weyl algebra ${\cal E}(\tilde{\cal W})$.

The correspondence between the Fedosov $\circ$-product and the Moyal $*$-product of
Eq. \moyal\ is given when the involved fields $\phi^I(\tilde{\sigma},y,\zeta_B)$ are
{\it central}. That means that

\eqn\central{[ \tilde{\phi}^I, \tilde{\phi}^J]_B \equiv {1 \over i \zeta_B}
\bigg(\tilde{\phi}^I \buildrel{\circ}\over
{\wedge} \tilde{\phi}^J - (-1)^{q_1q_2} \tilde{\phi}^J
\buildrel{\circ}\over{\wedge} \tilde{\phi}^I \bigg) = 0}
for all $\tilde{\phi}^J \in {\cal E} \big ( \tilde W \otimes
\Lambda^{q_1}\big)$ and $\tilde{\phi}^J \in {\cal E} \big( \tilde{\cal W} \otimes
\Lambda^{q_2} \big)$, where $q_1$ and $q_2$ are the degrees of  $\tilde{\phi}^I$ and
$\tilde{\phi}^J$ as differentiable forms respectively. The set of central forms
is denoted by ${\cal Z} \otimes \Lambda$. Here ${\cal Z}$ coincides
with the algebra of quantum observables ${\cal M}$.
Let $\tilde{\phi}^I (\tilde{\sigma}, y, \zeta_B)$ be an element of
${\cal E} (\tilde {\cal W})$, we
define the symbol map $\sigma: {\cal E} (\tilde {\cal W}) \to {\cal Z},$ given by
\ $\tilde{\phi}^I (\tilde{\sigma}, y, \zeta_B) \mapsto \tilde{\phi}^I
(\tilde{\sigma},0, \zeta_B)$,
{\it i.e.} the map $\sigma$ is the
projection of ${\cal E}(\tilde{\cal W})$ onto ${\cal Z}$. Fields $\tilde{\phi}^I(\tilde
{\sigma},0,\zeta_B)$ are exactly the generic fields $\phi^I(\tilde{\sigma})$ of Sec. 2.2.

\bigskip
\noindent{\it Differential Operators}

One can define some important differential operators. The operator $
\delta:  {\cal E} \big ( \tilde {\cal W}_p \otimes \Lambda^q\big) \to
{\cal E} \big( \tilde {\cal W}_{p-1} \otimes \Lambda^{q+1}\big)$ defined
by $ \delta a \equiv d x^k \wedge {\partial a\over \partial y^k}$
and its dual operator $\delta^{\circ}: {\cal E} \big( \tilde {\cal W}_p
\otimes \Lambda^q \big) \to {\cal E} \big( \tilde {\cal W}_{p+1} \otimes
\Lambda^{q-1}\big) $ defined by $ \delta^{\circ}a \equiv y^k
{\partial \over \partial x^k} \rfloor a$
for all $a \in {\cal E} \big( \tilde{\cal W}_p \otimes \Lambda^q\big),$
where $\rfloor$ stands for the contraction. The operators $\delta$ and
$\delta^{\circ}$ satisfy several properties
very similar to those for the usual differential and co-differential; for
instance, there exists an analogue of Hodge-de Rham decomposition theorem \fedosov.

\bigskip
\noindent{\it Symplectic Connection}

Assume the existence of a torsion-free connection on $X$ which
preserves its symplectic structure. This connection is known as {\it
symplectic connection} ${\nabla}_i$.
This operator is a connection defined in the bundle $\tilde{\cal
W}$ as $ \nabla: {\cal E} \big( \tilde{\cal W} \otimes \Lambda^q\big)
\to {\cal E} \big( \tilde{\cal W} \otimes \Lambda^{q+1}\big) $ and is
defined in terms of the symplectic connection as $ \nabla a \equiv d x^i
\wedge \nabla_i a.$ In Darboux
local coordinates this connection is written as $ \nabla a = d a +
[\Gamma , a]_B$
where $\Gamma = {1\over 2} \Gamma_{ijk} y^i y^j d x^k$ is a local one-form
with values in ${\cal E} \big( \tilde {\cal W} \big),$ $\Gamma_{ijk}$ are
the symplectic connection's coefficients, $d = d x^i \wedge {\partial
\over \partial x^i}$ and $\nabla_i$ is the covariant derivative on
$X$ with respect to ${\partial \over \partial x^i}.$

Following Fedosov, we define a more general connection $D$ in the Weyl
bundle $\tilde{\cal W}$ as follows
\eqn\conn{ Da = \nabla a +
[\gamma,a]_B,}
where
$\gamma \in {\cal E}({\cal E}(\tilde{\cal W}) \otimes \Lambda^1)$ is globally
defined on $X$. The {\it curvature} of the connection $D$ is given
by $ \Omega = (R + \nabla \gamma + {1
\over i \hbar} \gamma^2),$
with the normalizing condition $\gamma_0 = 0$. Here $R$ is defined by $R:=
{1 \over 4} R_{ijkl}y^iy^j dx^k \wedge dx^l$ where $R_{ijkl}$ is the
curvature tensor of the symplectic connection. It can be shown that
for any section $a \in {\cal E}(\tilde{\cal W}\otimes \Lambda)$ one has
$D^2a =  [\Omega, a]_B. $

$X$ has two curvatures, one as Riemannian manifold $\tilde{R}_{ijlk},$ which is
defined by the linear connection on the tangent bundle of $X$ and as Fedosov manifold
the symplectic connection on the Weyl bundle has curvature $R_{ijkl}$. These
curvatures are related in the form \gelfand\

\eqn\curvature{\tilde{R}_{ijkl} = g_{ip} R^{p}_{ \ jkl} = g_{ip}\omega^{pm}R_{mjkl}.}
where $g_{ij}$ is the Riemannian metric on $X$.

Since the torus is flat, Riemann curvature is thus zero $\tilde{R}_{ijkl}
=0$ and consequently the curvature $R_{ijkl}$ of the symplectic connection
vanishes. Something similar occurs with the coefficients of the symplectic
connection $\Gamma_{ijk}$ since they are related to Christoffel symbols of
Riemannian geometry \gelfand. Thus $D$ is purely gauge and it is given by

\eqn\derivada{ D\tilde{\phi}^I = d \tilde{\phi}^I + [\gamma, \tilde
{\phi}^I]_B}
and the curvature $\Omega$ of $D$ is given by

\eqn\curva{ \Omega = d\gamma + {1 \over i \zeta_B} \gamma \buildrel{\circ}\over
{\wedge} \gamma.}
We will see later that $\gamma$  can be identified with the gauge fields.

\bigskip
\noindent {\it Abelian Connection}

One very important definition is that of the {\it Abelian connection}. A
connection $D$ is Abelian if for any section $a \in {\cal E}(\tilde{\cal
W} \otimes \Lambda)$ one has $ D^2 a = [ \Omega, a]_B = 0.$
From Eq. \central\ one immediately sees that the curvature $\Omega$ of the Abelian
connection is central. In Fedosov's paper the Abelian connection takes the form

\eqn\once{D = -\delta + \nabla + [r, \cdot],}
where $\nabla$ is a fixed symplectic connection and $r \in {\cal
E}(\tilde{\cal W}_3 \otimes \Lambda^1)$ is a globally defined one-form with
the Weyl normalizing condition $r_0 = 0$. This connection has curvature

\eqn\doce{ \Omega = - {1 \over 2} \omega_{ij} dx^i \wedge dx^j + R - \delta r +
\nabla r + { 1\over i \hbar} r^2 }
with $\delta r = R + \nabla r + {1 \over i \hbar} r^2. $
This last equation has  a {\it unique} solution satisfying the condition
$ \delta^{-1} r = 0.$

\bigskip
\noindent{\it Algebra of Quantum Observables}

Now consider the subalgebra ${\cal E}(\tilde{\cal W}_D)$ of ${\cal
E}(\tilde{\cal W})$ consisting of {\it flat sections} {\it i.e.}
$ {\cal E}(\tilde{\cal W}_D) = \{a\in {\cal E}(\tilde{\cal W}) | Da =
0\}.$
This subalgebra is called {\it the algebra of Quantum Observables}.

Now an important theorem  \fedosov\ is: For any $a_0 \in {\cal Z}$ there
exists a unique section $a\in {\cal E}(\tilde{\cal W}_D)$ such that
$\sigma(a) = a_0$.

As a direct consequence of this theorem we can construct a section $a \in
{\cal E}(\tilde{\cal W}_D)$ by its symbol $a_0 = \sigma(a)$ in the form

\eqn\exp{a = a_0 + \nabla_i a_0 y^i + {1 \over 2} \nabla_i \nabla_j a_0 y^i
y^j + {1 \over 6} \nabla_i \nabla_j \nabla_k a_0 y^i y^j y^k - {1
\over 24} R_{ijkl} \omega^{lm} \nabla_m a_0 y^i y^j y^k + ... \ .}

The last theorem states that there exists the bijective map
$\sigma: {\cal E}(\tilde{\cal W}_D) \to {\cal Z}.$
Therefore there exists the inverse map $\sigma^{-1}: {\cal Z} \to {\cal
E}(\tilde{\cal W}_D).$ It is possible to use this bijective map to recover the
Moyal product $*$ in ${\cal Z},$ $ a_0 * b_0 = \sigma (\sigma^{-1}(a_0) \circ
\sigma^{-1}(b_0)).$

In the case of the torus it is flat and Eq. \exp\ can be written as

\eqn\trece{\tilde{\phi}^I = \sum_{k=0}^{\infty} {1 \over k!}
( \partial_{i_1} \partial_{i_2}
... \partial_{i_k} \tilde{\phi}^I_0) y^{i_1} y^{i_2} ...y^{i_k}}
where $\tilde{\phi}^I_0$ is equal to the generic fields  $\phi^I$ of Sec. 2.
Flat sections on the Weyl bundle on $\tilde{\bf T}^p$ are described in terms of
the space of closed differential forms ${\cal E}(\tilde{\cal W}_D) = Ker \ D$. Thus
``physical states'' are  Moyal algebra-valued cohomology classes in
$H^*({\cal E}(\tilde{\cal W}_D))$ of the algebra of quantum observables
${\cal E}(\tilde{\cal W}_D).$

\bigskip
\noindent {\it Trace on the Weyl Algebra}

In order to work with a variational principle which involves Fedosov
geometry we would like to get a definition of trace. In the case $X
= {\bf R}^{2n}$ with the standard symplectic structure
$ \omega = \sum_i d p_i \wedge d q_i,$ the Abelian connection $D$ in
the Weyl bundle is $ - \delta + d$.
In this case the product $\circ$ coincides with the usual Moyal $*$-product
\fedosov. The trace in the Weyl algebra ${\cal E}(\tilde{\cal W}_D)$ over $ {\bf
R}^{2n}$ is the linear functional on the ideal ${\cal E}({\cal W}_D^{\rm
Comp})$ over ${\bf R}^{2n}$ (which consists of the flat
sections with compact support) given by

\eqn\traza{\tilde{tr} (a) =  \int_{X} \sigma (a)
{\omega^n \over n!}}
where $\sigma(a)$ means the projection on the center
$\sigma(a(x,y,\hbar)) := a(x,0,\hbar).$
This definition of the trace satisfies a series of useful properties
$i)$-. $tr (a \circ b) = tr (b \circ a)$
 and $ii)$-. $ tr (b) = tr (A_f b)$
for all the sections $a \in {\cal E}(\tilde{\cal W}_D)$, $b \in {\cal
E}(\tilde{\cal W}_D^{comp})$ where  ${\cal E}(\tilde{\cal W}_D^{comp})$ is the
Weyl algebra of sections with compact support. In last equation, $A_f$ is an isomorphism
$A_f: {\cal E}(\tilde{\cal W}_D^{comp})({\cal O}) \to {\cal E}(\tilde{\cal
W}_D^{comp})(f({\cal O})),$ where $f$ is a symplectic diffeomorphism of ${\bf R}^{2n}.$
It is also possible to construct a trace on the algebra of sections
${\cal E}(\tilde{\cal W}_D)$ for arbitrary symplectic manifolds $X$ . This trace
satisfies properties $i)$ and $ii)$ but unfortunately it is too formal and does
not have an explicit form.

 A series of papers involving the application of the geometry
 of deformation quantization in the context of
integrable systems
\ref\strachan{ I.A.B. Strachan, ``A Geometry for Multidimensional
Integrable Systems'',  hep-th/9604142.},
self-dual gravity
\ref\compean{H. Garc\'{\i}a-Compe\'an, J.F. Pleba\'nski and M. Przanowski,
``Geometry Associated with Self-dual Yang-Mills and the Chiral Model Approaches to
Self-dual Gravity'', Acta Physica Pol. {B} {\bf 29} (1998) 549,
hep-th/9702046.} and ${\cal W}$-gravity
\ref\castro{C. Castro, ``${\cal W}$ Geometry From Fedosov's Deformation
Quantization'', hep-th/9802023.}, are found in the literature.

\vskip 1truecm
\subsec{ Gauge Theory on Fedosov Manifolds}

In the last subsection we have reviewed some relevant facts of Fedosov's
deformation quantization theory, which we  now apply to the super Yang-Mills
theory coming from the matrix compactifications. We first  study the
general case of the gauge theory on a symplectic real manifold $X$ and its
description in terms of Fedosov geometry. After that we will show how to recover the
original theory with constant $B$-field on the dual torus worked out in Sec. 2.
Deformation quantization allows also the description of Yang-Mills theories
on K\"ahler spaces such as ALE and $K3$ spaces and Calabi-Yau threefolds.

\bigskip
\noindent{\it The General Case: Fedosov Manifolds}

We are going to study the general case of non-constant background $B$-field on the
general Fedosov manifold $X$ with the Riemannian structure $g_{ij}$. This is
described by a
supersymmetric Yang-Mills theory on the curved space $X$. As we have seen before
in Sec. 3.2,
the Moyal
algebra \moyal\ has to be substituted by the Weyl algebra on the tangent bundle to $X$.
Thus a field theory on a curved space parametrized by its symplectic structure enters
in the analysis.
${\cal M}$-valued gauge fields $A_{\mu}(\tilde{\sigma})$ on ${\bf R} \times
\tilde{\bf T}^p$ are promoted to
${\cal E}(\tilde{\cal W}_D)$-valued
connection gauge fields $\tilde{A}_{\mu}$ on ${\bf R} \times X$ given by

\eqn\gauge{ \tilde{A} = \sum_{\mu} \tilde{A}_{\mu}(x,y,\zeta_B) dx^{\mu} }
where  $x=x^{\mu} = (x^0,x^i)$, $x^0 =t$ and $x^i$ are the coordinates on
$X$. The field components $\tilde{A}_{\mu}$ are written as

\eqn\ilis{\tilde{ A}_{\mu}(x,y, \zeta_B) = \sum_{k =0}^{\infty} {1\over k!}
\big(\nabla_{i_1} \nabla_{i_2} \cdots \nabla_{i_k} {A}_{\mu } \big)
y^{i_1} y^{i_2} \cdots y^{i_k} - {1 \over 24} R_{ijkl}\omega^{lm} \nabla_m
A_{\mu} + \dots }
where $A_{\mu}=A_{\mu}(x, \zeta_B) = \tilde{A}(x,0,\zeta_B).$

The curvature of the gauge connection is given in geometrical terms as
$\tilde{F} = \nabla \tilde{A} + { 1\over i \zeta_B}
\tilde{A}\buildrel{\circ}\over{\wedge}\tilde{A}$.
In terms of the commutator it yields

\eqn\paty{\tilde{F} = \nabla \tilde{A} + [\tilde{A},\tilde{A}]_B,}
where
$\tilde{A} \in {\cal E}({\cal E}(\tilde{\cal W}_D) \otimes \Lambda ^1),$
$\tilde{F} \in {\cal E}({\cal E}(\tilde{\cal W}_D)\otimes \Lambda^2)$
and $\nabla$ is the symplectic connection.

The scalar fields are expressed by

\eqn\lucy{\tilde{ X}^A = \sum_{k =0}^{\infty} {1\over k!}
\big(\nabla_{i_1} \nabla_{i_2} \cdots \nabla_{i_k} {X}^A \big)
y^{i_1} y^{i_2} \cdots y^{i_k} - {1 \over 24} R_{ijkl}\omega^{lm}\nabla_m X^A +
\dots}
and their interaction with the gauge fields $\tilde{A}_{\mu}$ are given by the
covariant derivative
$D\tilde{ X}^A = \nabla\tilde{X}^A + {1 \over i \zeta_B}\tilde{A}\buildrel{\circ}\over{\wedge}
\tilde{ X}^A$. In terms of the commutator it yields
\eqn\alej{ D\tilde{ X}^A = \nabla\tilde{X}^A +  [\tilde{A},\tilde{ X}^A]_B}
where $D$ is the connection in the Weyl bundle $\tilde{W}.$ Here we have
identified the global one-form $\gamma$ with the gauge fields $\tilde{A}_{\mu}.$
The Lagrangian \lagrangiano\ can be generalized to the Weyl bundle on a curved
compact space $X$ as follows

$$
L_M = \tilde{\rm Tr}
 \bigg( - {1 \over 4 T^2} \tilde{F}_{\mu \nu} \circ \tilde{F}^{\mu \nu} -
{1 \over 2 } g^{\mu \nu}D_{\mu} \tilde{X}^A \circ D_{\nu} \tilde{X}^A $$

\eqn\carmen{ + [\tilde{X}^A , \tilde{X}^B]_B \circ [\tilde{X}^A , \tilde{X}^B]_B
-{i \over 2} \tilde{\Theta}^{\alpha} \circ \Gamma^{\mu}_{\alpha \beta}{D}_{\mu}
\tilde{\Theta}^{\beta} +
{T\over 2} \tilde{\Theta}^{\alpha} \circ
\Gamma^A_{\alpha \beta} [\tilde{X}_A ,\tilde{\Theta}^{\beta}]_B \bigg)}
where $g_{\mu \nu}$ is the Riemannian metric on ${\bf R} \times X$ and
$\tilde{\rm Tr}$ consists of the Fedosov's trace for the general case and the
matrix trace
Tr$_N$. General Fedosov's trace
does exist but an explicit form is not known \fedosov.  In the most general
case the spinors
$\tilde{\Theta}^{\alpha}$ are symplectic spinors on Fedosov
manifolds. They are well defined if a metaplectic structure is added on $X$ as
was shown in \ref\habermann{K. Habermann, ``Basic Properties of Symplectic
Dirac Operators'', Commun. Math. Phys. {\bf 184} (1997) 629.}. Covariant derivative
in general involves the gauge and symplectic connections

\eqn\general{ D_{\mu} \tilde{\varphi}^A = \partial_{\mu} \tilde{\varphi}^A +
[\tilde{A}_{\mu} +
 \Gamma_{\mu}, \tilde{\varphi}^A]_B}
where $\tilde{\varphi}^A$ stands for the fields $\tilde{X}^A$ and $\tilde{\Theta}^{\alpha}$.
Lagrangian  \carmen\ corresponds to a gauge theory with Weyl algebra-valued fields in the curved
space ${\bf R}\times X$ and represents deformations of a gauge theory on a curved space
parametrized by a symplectic structure in $X$ given by the $B$-field.

Therefore we have been able to find an alternative way of obtaining the deformed
gauge theory. This deformation was suggested in \connes\ where was argued that
it could be obtained following the paper \alanpaper. Lagrangian \carmen\
generalizes  \lagrangiano\ and we shall show how to recover the latter form the former.
This deformed gauge theory has been the natural generalization to Fedosov
deformation quantization theory to define the procedure of quantization
in a global way. This involves curved manifolds and so the gauge theory becomes
defined on these spaces. Automatically is obtained the generalization to non-constant
$B$-fields on the underlying compact space. Thus the effect of the $B$-field
in matrix compactification on a curved space $X$ is the introduction of a symplectic
structure on $X$ compatible with the Riemannian structure and thus deformation
quantization geometry is very important to study these compactifications.

There is also some relation with the
work \volovich, where matrix theory on a curved space is defined as the dimensional
reduction the 10-dimensional Yang-Mills on curved 10-dimensional manifold $Y$. The
relation with this work is given if we identify $Y$ with $X \times {\bf R}^{10 -p}$
and the corresponding metric $g_{IJ}$ on $Y$ with the metric on $X$ given by
the Riemannian metric $g_{ij}$ and on ${\bf R}^{10-p}$ the flat metric $\eta_{AB}$.
Our results might also be related to other works of matrix theory on curved spaces
\refs{\ooguri,\curved}. Perhaps a preliminary step in seeking this relation
 would be first finding a relation of our results to the paper \curved.
In the general case supersymmetry would be completely broken or a part of it
depending on the geometric structure of $X$. If $X$ is a K\"ahler manifold the
preserved supersymmetry is ${\cal N}=1,$ $d=4$ and for a hyperk\"ahler is
${\cal N}=2,$ $d=4$.

 Finally but not least
important is that the deformed gauge theory is also a non-local field theory
which generalizes the non-local theories studied in
\berkooz. The renormalization group behavior of these theories is much more involved
and deserves a careful study which we will take up in a future work.

\bigskip
\noindent
{\it Gauge Theory on the Torus}

 Now we will show how to recover Lagrangian \lagrangiano\
 from the general theory described by
 the Lagrangian \carmen. First of all we  set $X=\tilde{\bf T}^p$ and repeat the
 analysis of the last subsection. Thus the gauge
 fields can be written as

\eqn\norma{ \tilde{A} = \sum_{\mu} \tilde{A}_{\mu}(\tilde{\sigma},y,\zeta_B)
d\tilde{\sigma}^{\mu} }
where
\eqn\lina{\tilde{ A}_{\mu}(\tilde{\sigma},y,\zeta_B) = \sum_{k =0}^{\infty} {1\over k!}
\big(\partial_{i_1} \partial_{i_2} \cdots \partial_{i_k} A_{\mu} \big)
y^{i_1} y^{i_2} \cdots y^{i_k}}
where $A_{\mu}$ corresponds precisely to the gauge field which enters in
Lagrangian \lagrangiano. Also the symplectic connection is substituted
by partial derivatives because the
coefficients of the connection $\Gamma$ for the torus vanish.

The curvature of the gauge connection $\tilde{F} = d \tilde{A} + { 1 \over i \zeta_B}
 \tilde{A}
\buildrel{\circ}\over{\wedge}\tilde{A}$.
In terms of the commutator it yields

\eqn\karina{\tilde{F} = d\tilde{A} + [\tilde{A},\tilde{A}]_B,}
where
$\tilde{A} \in {\cal E}({\cal E}(\tilde{\cal W}_D) \otimes \Lambda ^1)$
and $\tilde{F} \in {\cal E}({\cal E}(\tilde{\cal W}_D)\otimes \Lambda^2)$.
In components it can be written as

\eqn\components{\tilde{F}_{\mu \nu} = \partial_{\mu}\tilde{A}_{\nu}
- \partial_{\nu}\tilde{A}_{\mu} + [\tilde{A}_{\mu},\tilde{A}_{\nu}]_B,}

The scalar fields are

\eqn\linda{\tilde{ X}^A = \sum_{k =0}^{\infty} {1\over k!}
\big(\partial_{i_1} \partial_{i_2} \cdots \partial_{i_k} {X}^I \big)
y^{i_1} y^{i_2} \cdots y^{i_k}}
and the interaction with the gauge fields are given by the covariant derivative
$D\tilde{ X}^A = d \tilde{X}^A + {1 \over i \zeta_B}\tilde{A}\buildrel{\circ}\over{\wedge}
\tilde{ X}^A$ or in terms of the commutator
\
\eqn\monica{ D\tilde{\bf X}^A = d\tilde{\bf X}^A + [\tilde{A},\tilde{\bf X}^A]_B.}
Substituting these expressions into the general Lagrangian  \carmen\ we get

$$
L_M = \tilde{\rm Tr}
 \bigg( - {1 \over 4 T^2} \tilde{F}_{\mu \nu} \circ \tilde{F}^{\mu \nu} -
{1 \over 2 } \eta^{\mu \nu} D_{\mu} \tilde{X}^A \circ D_{\nu}\tilde{X}^A $$

\eqn\fin{ + [\tilde{X}^A , \tilde{X}^B]_B \circ [\tilde{X}^A , \tilde{X}^B]_B
-{i \over 2}\tilde{\Theta}^{\alpha} \circ \Gamma^{\mu}_{\alpha \beta}D_{\mu}
\tilde{\Theta}^{\beta} +
{T\over 2} \tilde{\Theta}^{\alpha} \circ
\Gamma^A_{\alpha \beta} [\tilde{X}_A, \tilde{\Theta}^{\beta}]_B \bigg)}
where
\eqn\usual{ D_{\mu}\tilde{X}^A = \partial_{\mu}\tilde{X}^A +
 [\tilde{A}_{\mu}, \tilde{X}^A]_B}
and
\eqn\nousual{ D_{\mu}\tilde{\Theta}^{\alpha} = \partial_{\mu}\tilde{\Theta}^{\alpha} +
[\tilde{A}_{\mu}, \tilde{\Theta}^{\alpha}]_B.}

In this case we have a flat geometry  and Fedosov trace \traza\ is valid. then
substituting \fin\  into the last Lagrangian it yields

$$
L_M = \int_{\tilde{\bf T}^p}  {\omega^{p/2} \over ({p\over 2})!}\  {\rm Tr}_N \
\sigma
\bigg( - {1 \over 4 T^2} \tilde{F}_{\mu \nu} \circ \tilde{F}^{\mu \nu} -
{1 \over 2 } \eta^{\mu \nu} D_{\mu} \tilde{X}^A \circ D_{\nu}\tilde{X}^A $$

\eqn\finfedov{ + [\tilde{X}^A , \tilde{X}^B]_B \circ [\tilde{X}^A , \tilde{X}^B]_B
-{i \over 2}\tilde{\Theta}^{\alpha} \circ  \Gamma^{\mu}_{\alpha \beta}D_{\mu}
\tilde{\Theta}^{\beta} +
{T\over 2} \tilde{\Theta}^{\alpha} \circ
\Gamma^A_{\alpha \beta} [\tilde{X}_A ,\tilde{\Theta}^{\beta}]_B \bigg)}
where $\sigma$ is the isomorphism or symbol map which is a projection
of ${\cal E}(\tilde{\cal W})$ into ${\cal M}_B$ arising in \traza. Using the fact that
$\omega^{p/2}$ is
proportional to the volume form of the dual torus, $d^p \tilde{\sigma}$ then we obtain
precisely Lagrangian \lagrangiano\ for the ${\cal M}$-valued fields. Thus we have
shown that
general Lagrangian constitutes its natural generalization from the point of view of
deformation quantization theory and it contains the gauge theory \lagrangiano,
derived from matrix theory, as a particular case.

\bigskip
\noindent{\it Gauge Theory on K\"ahler-Fedosov Manifolds }

Deformed gauge theory is also well defined  when $X$ is a K\"ahler manifold. A
Fedosov $*$-product of the Wick type can be defined \ref\bordemann{M. Bordemann and
S. Waldmann, ``A Fedosov Star Product of Wick
Type for K\"ahler Manifolds'', q-alg/9605012.}.

Let $X$ be a K\"ahler manifold with local coordinates $(z_1, \dots ,z_n,
\bar{z}_1, \dots ,\bar{z}_n)$. The symplectic form is given by $\omega
= {1 \over 2} B_{i \bar{j}} dz^i \wedge d z^{\bar{j}}$ with $B_{i \bar{j}}$ positive
definite. The Poisson tensor is given by $\Lambda = {2 \over i} B^{i \bar{j}} Z_i \wedge
\bar{Z}_{\bar{j}}$ with $Z_i = {\partial \over \partial z_i}$ and
$\bar{Z}_{\bar{j}} = {\partial \over \partial \bar{z}_{\bar{j}}}$. A Fedosov product
on the sections of the Weyl bundle on $X$ can be defined as

\eqn\luz{ \tilde{\phi}^I \circ ' \tilde{\phi}^J = \sum_{r=0}^{\infty}
({i \zeta_B \over 2})^r M_r(\tilde{\phi}^I,\tilde{\phi}^J)}
where
\eqn\amorina{M_r(\tilde{\phi}^I,\tilde{\phi}^J) = {1 \over r!} ({4 \over i})^r
B^{i_1 \bar{j}_1} \dots B^{i_r \bar{j}_r} i_s(Z_{i_1}) \dots i_s(Z_{i_r})
\tilde{\phi}^I  i_s(\bar{Z}_{\bar{j}_1}) \dots i_s(\bar{Z}_{\bar{j}_r})
\tilde{\phi}^J.}
Here $i_s(Z_{i_r})\tilde{\phi}^I$ is the insertion (symmetric substitution) of
the vector field $Z_{i_r}$ in the symmetric part of $\tilde{\phi}^I$.

With this product a similar analysis of Subsec. 3.2 and 3.3 can be done with
this holomorphic
Fedosov product. In this context the spinors are well defined and can be
treated following the formalism of \nref\katha{
K. Habermann, ``Symplectic Dirac Operators on K\"ahler Manifolds'', Preprint
1998.} \katha.

In particular analysis of $K3$ and Calabi-Yau manifolds is
then possible. We leave the application of deformation quantization on K\"ahler
spaces to matrix compactifications on ALE and $K3$ spaces and also to Calabi-Yau
threefolds for future research.

\vskip 2truecm
\newsec{Concluding Remarks}

In this paper we have shown that the description of matrix compactifications on
noncommutative tori corresponds to a genuine Weyl correspondence of deformation
quantization theory in the terms of \bayen. This description leads to a natural
generalization from the classical Yang-Mills theory on the noncommutative torus to the
consideration of
curved spaces through the Fedosov's geometry of deformation quantization theory. In the
description on curved spaces we have gained automatically the generalization
to non-constant background $B$ fields on the curved manifolds.
However the application of the full formalism of deformation
quantization to matrix theory still remains to be done. It is interesting to note that
Fedosov
deformation quantization is well defined on general K\"ahler manifolds
\bordemann, in particular,
on ALE spaces, $K3$ spaces and Calabi-Yau threefolds. It would be interesting to
investigate whether
deformation quantization sheeds some light in ALE, $K3$ and Calabi-Yau
compactifications of matrix theory in a background $B$-field.

Lagrangian \lagrangiano\ is  generalized to flat connections on the Weyl bundle
on a general symplectic and Riemannian manifold $X$. In the
case of toroidal compactifications this description is rather trivial and Weyl bundle
description is equivalent to the gauge theory described in Sec. 2.3. Some contact with
a definition of matrix theory on curved spaces \volovich\ has been established.
 This would be
useful to make contact of our results with other definitions of matrix theory
on noncommutative curved spaces  \refs{\ooguri,\curved}.
It is known that obstructions to get a global $*$-product on arbitrary
symplectic manifolds may arise as
global anomalies \baadhio. It would be  interesting to check whether the
general Lagrangian \carmen\
is global anomaly free. Finally, very recently in \nref\mreuter{M. Reuter, ``Quantum
Mechanics as a Gauge Theory of Mataplectic Spinor Fields'', to appear in Int. J.
Mod. Phys. A, hep-th/9804036.} Ref.\mreuter, a relation between quantum mechanics and
Yang-Mills theory has been found. This relation contains several ingredients and
constructions which seems to coincide with ours in some points. It would be very
interesting to find a relation between both approaches.

\vskip 2truecm


\centerline{\bf Acknowledgments}
I would like to thank A. G\"uijosa, P-M. Ho,  J. Pleba\'nski, M. Przanowski,
 A. Uranga and Y-S. Wu  for useful
discussions and comments, K. Habermann for send me her paper \katha\ and M. Reuter
for useful correspondence.
This work is supported by a Postdoctoral
CONACyT fellowship under the program {\it Programa de
Posdoctorantes: Estancias Posdoctorales en el Extranjero para Graduados en
Instituciones Nacionales 1996-1997}. Thanks are due  E. Witten and
the Institute for Advanced Study for its hospitality.

\listrefs
\end